\documentclass[11pt]{article} 
\usepackage[utf8]{inputenc} 

\usepackage{geometry} 
\usepackage{graphicx}

\usepackage{booktabs}
\usepackage{array} 
\usepackage{paralist} 
\usepackage{verbatim} 
\usepackage{subfig} 

\usepackage{fancyhdr}
\usepackage{sectsty}
\allsectionsfont{\sffamily\mdseries\upshape} 

\usepackage[nottoc,notlof,notlot]{tocbibind} 
\usepackage[titles,subfigure]{tocloft}

\title{Geometrisation of Fermions in Flat Spacetimes}
\author{William J. Leigh\thanks{wl197@leicester.ac.uk} \\Division of Library and Learning Services\\University of Leicester, University Road, Leicester LE1 7RH\\United Kingdom}
\date{March 2024}

\begin{document}
\maketitle

%\begin{center}{\bf{Statements and Declarations}}\\The author declares that no external financial or other interests\\exist in connection with the work reported in this paper.\end{center}

\begin{abstract}Requiring physical consistency in a classical flat spacetime geometrisation of fermions is shown to suggest the introduction of torsion. A resulting simple model for that torsion produces a localised quantum-like particle as a solution of a spinor identity that closely resembles the Dirac equation of quantum electrodynamics. All relevant integrals involving solutions of the spinor identity converge and require that the spin, which is no longer intrinsic but arises from circulating currents, be 1/2 whilst the magnetic moment takes the quantum value of 1 magneton, characteristic of an isolated Dirac particle. The underlying torsion associates the otherwise strictly localised particle with an extended spacetime structure that may be relevant to wider quantum phenomena.  \end{abstract}

{\bf{Keywords}} geometrisation, spinor, torsion, spacetime, fermion

\section{Introduction}

This is the first of two formulational papers together intended to provide the basis for an account of fermions in which geometric properties of spacetime play a crucial role in combining classical and quantum features of these particles. The (–3,1) metric signature of Maxwellian electrodynamics ensures the purely mathematical existence of classical spinor fields, geometric objects that underlie the current work, while the standard scale of quantum electrodynamics [QED] is anticipated by using units for which ${\hbar = c = 1}$.

In contrast with the way they appear in QED, fermions here turn out not to correspond to hypothetical classical point particles but have instead a spatial extension that is effectively determined by the quantum Compton wavelength. The spatially decaying exponents this entails guarantee convergence of integrals and yield what are normally understood to be specifically quantum properties for the elementary fermion, namely the spin of 1/2 and magnetic moment of 1 magneton that are characteristic of a free Dirac particle.

The current paper treats only flat spacetimes, that is to say, ones for which the Riemann tensor vanishes. The simplest of these, deriving its physical significance from Maxwellian electromagnetism, will be denoted {\em Poincaré spacetime} after the group of translations and generalised rotations under which it and its Minkowski metric are invariant. 

The essential physical input in this paper, apart from Maxwell’s theory, is the insistence in Section~\ref{sec:poincare} that there exist a unique physical spinor field accounting for the presence of a stationary fermion at the origin of the space coordinate system. Given the purely geometric relationships existing between spinor and vector fields, such a spinor field together with its putative electromagnetic potential must satisfy a certain differential equation that is referred to here as the {\em spinor identity} and resembles, but is not equivalent to, the electromagnetically-coupled Dirac equation of QED.%\footnote{\cite{badejehle} See Section X, p.727.}.

It transpires however towards the end of Section~\ref{sec:poincare} that the constraints needed to obtain physically acceptable solutions can only be satisfied in an empty Poincaré spacetime. Nevertheless, the way in which non-trivial currents associated with the physical spinor and its electromagnetic field fail to be reconciled suggests a natural extension of Poincaré geometry which, maintaining the Minkowski metric, incorporates a torsion that may be associated with entanglement of localised fermions. Section~\ref{sec:cochlean} defines the corresponding {\em  Cochlean spacetime} and identifies the differential equation determining the form of the physical spinor. Simple solutions are presented and classified as ‘moderated’ and ‘unmoderated’, according to the nature of the exponential spatial decay that assures particle-like localisation.
 
After deriving a form for the moderated electromagnetic potential by identifying spinor and electromagnetic currents, Section~\ref{sec:physics} hypothesises an energy density tensor for the fermion field. A total energy-momentum tensor is then constructed and used to develop an analysis of the mass, spin, charge and magnetic moment of states corresponding to simple moderated solutions of the spinor identity. The properties established for these solutions lead to the well-known results of relativistic quantum mechanics for essentially free particles. 

A realistic account of a fermion in Cochlean spacetime can be expected to require a more complex, composite solution to the spinor identity. Given the primacy of geometry in the current approach however and the need therefore to extend this work to curved spacetimes, calculations using such solutions are not provided in the present paper, in keeping with its purely formulational character.

\section{The Poincaré Spinor Identity}~\label{sec:poincare}

The spinor notation of this and later sections is adapted from the review article by Bade and Jehle \cite{badejehle} as well as from the work of Penrose and Rindler \cite{penrose01,penrose02} and the underlying reasoning, although largely familiar, is expounded explicitly here to facilitate the presentation of its further development in Section~\ref{sec:cochlean}.

\subsection{Tensor-Spinor Relations and the Spinor Identity}

The 2-spinor Poincaré approach uses four complex 2x2 matrices, often known as Infeld-van-der-Waerden symbols [IvdW1], that are partially defined by the condition 
\begin{equation}
\sigma _A^{\mu \dot X}\sigma _{\dot X}^{\nu B} + \sigma _A^{\nu \dot X}\sigma _{\dot X}^{\mu B} =  - {g^{\mu \nu }}e_A^{\; \bullet B};\;e_A^{\; \bullet B} = \delta _A^{\;B}
\label{(2.1.1)},\end{equation}
where ${g^{\mu \nu }}$  is the symmetric Minkowski metric, and $e_{AB}$ the skew 2-spinor metric that lowers a spin index by postmultiplication, $e^{AB}$ raising by premultiplication.
Using IvdW1 to establish a correspondence between real 4-vectors and products of 2-spinors with their conjugates, as in say \begin{equation}{j_\mu } = \sigma _\mu ^{A\dot B}{\xi _A}{\xi _{\dot B}},\end{equation}
requires the IvdW1 with both upper [or both lower] spin indices to be hermitian, but there still remains considerable freedom in the choice of these matrices. A common choice, based upon the Pauli spin matrices for rectangular coordinates, leads to the forms in cylinder polar coordinates $\left( {\rho ,\varphi ,z,t} \right)$, for which ${g^{\mu \nu }} = \left( { - 1, - {\rho ^{ - 2}}, - 1,1} \right)$],

\begin{equation}\begin{array}{ll}
  \sigma _A^{1\dot B} = \frac{1}{{\sqrt 2 }}\left( {\begin{array}{*{20}{c}}
  { - {e^{i\phi }}}&0 \\ 
  0&{{e^{ - i\phi }}} 
\end{array}} \right);&\sigma _A^{3\dot B} = \frac{1}{{\sqrt 2 }}\left( {\begin{array}{*{20}{c}}
  0&1 \\ 
  1&0 
\end{array}} \right); \\ 
  \sigma _A^{2\dot B} = \frac{{ - i}}{{\sqrt 2 \rho }}\left( {\begin{array}{*{20}{c}}
  {{e^{i\phi }}}&0 \\ 
  0&{{e^{ - i\phi }}} 
\end{array}} \right);&\sigma _A^{4\dot B} = \frac{1}{{\sqrt 2 }}\left( {\begin{array}{*{20}{c}}
  0&{ - 1} \\ 
  1&0 
\end{array}} \right).  
\end{array}\label{(2.1.3)}\end{equation}

Cylinder polars will be used throughout the paper since they offer the most natural framework for describing a system spinning around a conventional z axis and centred at the origin of Poincaré space coordinates. Covariant derivatives will be used with these coordinates, with torsion-free Levi-Cività connections being defined in the usual way by requiring covariant constancy of the spacetime metric. Defining covariant tensor derivatives using ${\tilde d_\nu }{v_\mu } := {d_\nu }{v_\mu } - \tilde \gamma _{ \bullet \mu \nu }^\eta {v_\eta }$ the connections are then \[{\tilde \gamma _{\lambda \mu \nu }} = \frac{1}{2}\left( {{g_{\lambda \mu ,\nu }} + {g_{\lambda \nu ,\mu }} - {g_{\mu \nu ,\lambda }}} \right) = \rho \left( {{\delta _{\lambda \mu 12}}{\delta _{\nu 2}} - {\delta _{\lambda 2}}{\delta _{\mu 2}}{\delta _{\nu 1}}} \right);\;{\delta _{\lambda \mu 12}} := {\delta _{\lambda 1}}{\delta _{\mu 2}} - {\delta _{\lambda 2}}{\delta _{\mu 1}}.\]
It should be noted that the obvious covariant constancy of the Pauli matrices with respect to rectangular coordinates is of course preserved in the passage to cylinder coordinates, so that the conventional (\ref{(2.1.3)}) satisfy
\begin{equation}{\tilde d_\mu }\sigma _A^{\nu \dot B} = {\tilde d_\mu }\sigma _{A\dot B}^\nu  = {\tilde d_\mu }{\sigma ^{\nu A\dot B}} = 0.\label{(2.1.4)}\end{equation}

When Poincaré spinor connections are introduced then, noting that product pairs of spinors and conjugate spinors always occur in transformations between spinorial and tensorial quantities, it is reasonable to require \cite{penrose01} that such connections preserve the covariant constancy of the product of the spinor metric and its conjugate. Insisting further that Poincaré spinor connections be skew, their form is fixed by 
\[\begin{array}{l}
   \quad {D_\mu }\left( {{e_{AB}}{e_{\dot C\dot D}}} \right) = 0 \hfill \\
   \Leftrightarrow {d_\mu }\left( {{e_{AB}}{e_{\dot C\dot D}}} \right) - \gamma _{\mu A}^{\quad  \bullet X}{e_{XB}}{e_{\dot C\dot D}} - \gamma _{\mu B}^{\quad  \bullet X}{e_{AX}}{e_{\dot C\dot D}} - {e_{AB}}\gamma _{\mu \dot C}^{\quad  \bullet \dot Y}{e_{\dot Y\dot D}} - {e_{AB}}\gamma _{\mu \dot D}^{\quad  \bullet \dot Y}{e_{\dot C\dot Y}} = 0 \hfill \\
   \Leftrightarrow {\rm Re} {\gamma _{\mu \left[ {AB} \right]}} = 0 \Leftrightarrow {\gamma _{\mu AB}} =  - iq{a_\mu }{e_{AB}}. \hfill  
\end{array} \]
The real vector coefficient here in the spinor connection is written in a way intended to suggest its identification as a geometrised electromagnetic potential. As will shortly appear however, such a potential is incompatible with the standard QED identification of the spinor current as the source of electromagnetism.

Assuming then there exists a unique {\em physical spinor}, $\xi$, that describes a stable stationary fermion at the origin of space coordinates, the covariant derivative may be applied to it to define a secondary, conjugate spinor, $\zeta$, through
\begin{equation}D_{\dot A}^{ \bullet B}{\xi _B} = \sigma _{\dot A}^{\mu B}\left( {{d_\mu } + iq{a_\mu }} \right){\xi _B} =  - \frac{m}{{\sqrt 2 }}{\zeta _{\dot A}},\label{(2.1.5)}\end{equation}
where $m$, a real constant with the dimension of mass, has been introduced. Applying the same, conjugated derivative to the secondary spinor must, from the uniqueness assumption, give a constant multiple of the physical spinor; rescaling the secondary spinor and renaming the mass factor as needed allows one to derive the identity
\begin{equation}D_A^{ \bullet \dot B}{\zeta _{\dot B}} = \sigma _A^{\mu \dot B}\left( {{d_\mu } - iq{a_\mu }} \right){\zeta _{\dot B}} =  - \frac{m}{{\sqrt 2 }}{\xi _A}. \label{(2.1.6)}\end{equation}
That signs differ between these two identities shows the system to be inequivalent to the original 4-spinor minimally coupled first-order equation proposed by Dirac.\footnote{See Bade and Jehle \cite{badejehle}, (I 10).}

Applying the operator in (\ref{(2.1.6)}) to (\ref{(2.1.5)}), there follows the further, second-order identity
\begin{equation}D_A^{ \bullet \dot X}D_{\dot X}^{ \bullet B}{\xi _B} = \frac{{{m^2}}}{2}{\xi _A}.\label{(2.1.7)}\end{equation}
It is this form that is properly the subject of the current paper, and it will here be referred to as the {\em spinor identity}.\footnote{See Bade and Jehle also for the Schrödinger-Klein-Gordon equation. Note that the term ‘spinor identity’ has been judged preferable to coinages such as ‘pseudo/quasiDirac equation’, even though (\ref{(2.1.7)}) can only be a true ‘identity’ for the physical spinor, assuming such a spinor to exist.}  Whilst the form of this identity for the physical spinor suggests it is a mass shell condition, it would be premature to identify the mass parameter – as a QED bare mass, say – until and unless physical solutions to (\ref{(2.1.7)}) are available; see Section~\ref{sec:cochlean} below.

The LHS of the spinor identity can be expanded as follows
\begin{eqnarray}\label{(2.1.8)}
  D_A^{ \bullet \dot X}D_{\dot X}^{ \bullet B}{\xi _B} & = & \sigma _A^{\mu \dot X}{D_\mu }\left\{ {\sigma _{\dot X}^{\nu B}{D_\nu }{\xi _B}} \right\} = \sigma _A^{\mu \dot X}\left\{ {{D_\mu }\sigma _{\dot X}^{\nu B}} \right\}{D_\nu }{\xi _B} + \sigma _A^{\mu \dot X}\sigma _{\dot X}^{\nu B}\left\{ {{D_\mu }{D_\nu }{\xi _B}} \right\}\nonumber \\ 
   & = & \sigma _A^{\mu \dot X}\left\{ {{{\tilde d}_\mu }\sigma _{\dot X}^{\nu B} - 2iq{a_\mu }\sigma _{\dot X}^{\nu B}} \right\}{D_\nu }{\xi _B} + \left( {\sigma _A^{\mu \nu B} - \frac{1}{2}{g^{\mu \nu }}e_A^{ \bullet B}} \right){D_\mu }{D_\nu }{\xi _B}\nonumber \\ 
   & = & \left( {\sigma _A^{\mu \nu B} - \frac{1}{2}{g^{\mu \nu }}e_A^{ \bullet B}} \right)\left( {{D_\mu } - 2iq{a_\mu }} \right){D_\nu }{\xi _B} \\ 
   & = & \left( {\sigma _A^{\mu \nu B} - \frac{1}{2}{g^{\mu \nu }}e_A^{ \bullet B}} \right)\left( {{{\tilde d}_\mu } - iq{a_\mu }} \right)\left( {{d_\nu } + iq{a_\nu }} \right){\xi _B}\nonumber \\ 
   & = &  - \frac{1}{2}{g^{\mu \nu }}\left( {{{\tilde d}_\mu } - iq{a_\mu }} \right)\left( {{d_\nu } + iq{a_\nu }} \right){\xi _A} + 2iq\sigma _A^{\mu \nu B}{a_\nu }{d_\mu }{\xi _B} + i\frac{q}{2}{f_{\mu \nu }}\sigma _A^{\mu \nu B}{\xi _B},\nonumber
\end{eqnarray}
with symmetric-spinor/skew-tensor electromagnetic fields and IvdW2 defined by

\begin{equation}\label{(2.1.9)}{f_{AB}} := {f_{\mu \nu }}\sigma _{AB}^{\mu \nu };\;{f_{\mu \nu }} := {\tilde d_\mu }{a_\nu } - {\tilde d_\nu }{a_\mu } = {d_\mu }{a_\nu } - {d_\nu }{a_\mu };\;\sigma _{AB}^{\mu \nu } := \frac{1}{2}\left[ {\sigma _A^{\mu \dot X}\sigma _{\dot XB}^\nu  - \sigma _A^{\nu \dot X}\sigma _{\dot XB}^\mu } \right].\end{equation}
The spinor identity can then be written as an expression for the coupling term

\begin{equation}\label{(2.1.10)}iqf_A^B{\xi _B} = \left[ {{g^{\mu \nu }}\left( {{{\tilde d}_\mu } - iq{a_\mu }} \right)\left( {{d_\nu } + iq{a_\nu }} \right) + {m^2}} \right]{\xi _A} - 4iq\sigma _A^{\mu \nu B}{a_\nu }{d_\mu }{\xi _B},\end{equation}
a form differing from the second-order Dirac equation in the alternation of signs in the inner brackets and in the presence of the final term. The source of both features is the covariant derivative of the IvdW1 appearing in (\ref{(2.1.8)}), and provided a non-trivial spinor connection is used to geometrise the electromagnetic potential this derivative cannot be arranged to vanish, whether or not, as assumed here, the purely tensor part of the derivative of the IvdW1 is zero.

In considering what form exact solutions to (\ref{(2.1.7)}) should have it is natural to ask whether there exists a physical spinor that both exhibits the oscillatory time behaviour characteristic of a putative single particle quantum state, and is also consistent with the azimuthally uniform and steady electric current needed to produce a static and axially-symmetric electromagnetic fermion field. Against this background then, and assuming the freedom to combine Poincaré and gauge transformations, it will be supposed that the components of the physical spinor have been arranged to take the modulus-argument forms appearing in

\[{\xi _A} = H\left( {\begin{array}{*{20}{c}}
  {{e^{i\chi }}} \\ 
  {{e^{ - i\chi }}} 
\end{array}} \right),\quad \quad \chi  = \Theta  + \Phi ,\quad \Phi  := \phi /2 - mt\]
where $H $ and $\Theta $ are taken to be functions only of the $\rho $ and $z $ cylinder polar spatial coordinates. Corresponding to these conditions, the geometrised electromagnetic potential that appears in (\ref{(2.1.8)}) will be supposed to have non-vanishing azimuthal and time components only, and these components will themselves be supposed independent of azimuth and time.

Two conserved current vectors appear in this formulation: the first purely geometric, arising from the spinor connection that gives electromagnetism, the second physical, arising from spinor matter. The normal understanding of how charge and matter are related requires these two conserved currents to be multiples of each other. 

That the electromagnetic current is conserved is guaranteed by its definition as the divergence of the skew electromagnetic field from

\[j_\alpha ^e := {g^{\nu \beta }}{\tilde d_\nu }{f_{\alpha \beta }} \Rightarrow {g^{\mu \alpha }}{\tilde d_\mu }j_\alpha ^e = {g^{\mu \alpha }}{g^{\nu \beta }}{\tilde d_\mu }{\tilde d_\nu }{f_{\alpha \beta }} \equiv 0,\]
where, since a Poincaré spacetime has no torsion, the covariant derivatives commute and give the final identity. In cylinder coordinates, with a potential that has time and azimuthal components only, the components of the electric current defined here are 
\begin{equation}j_2^e =  - \rho \left\{ {{{\vec \nabla }^2} - \frac{1}{{{\rho ^2}}}} \right\}\frac{{{a_2}}}{\rho },\quad j_4^e =  - {\vec \nabla ^2}{a_4}.\label{(2.1.11)}\end{equation}

That the spinor matter current be conserved is less easy to arrange, and how this can be done will now be investigated.

\subsection{The Spinor Matter Current}

The standard QED spinor matter current is the sesquilinear real form 
\begin{equation}\label{(2.2.1)}j_\mu ^s := \sigma _\mu ^{A\dot B}\left\{ {{\xi _A}{\xi _{\dot B}} + {\zeta _A}{\zeta _{\dot B}}} \right\} = {\xi ^t}{\sigma _\mu }{\xi ^ * } + {\zeta ^t}{\sigma _\mu }{\zeta ^ * },\end{equation}
and is conserved as a consequence of the Dirac equation, together with the covariant constancy of the IvdW1. In the present analysis, the Dirac equation does not apply; nevertheless, noting the vanishing of relevant Levi-Cività connections
\[\tilde \gamma _{ \bullet 22}^2 = \tilde \gamma _{ \bullet 22}^4 = \tilde \gamma _{ \bullet 24}^2 = \tilde \gamma _{ \bullet 24}^4 = 0,\]
the spinor current will still be conserved if it can be arranged to have only azimuthal and time components that are non-vanishing, with each being independent of both azimuth and time. These requirements pose an immediate problem however, as can be seen by noting that the hermitian IvdW1 in (\ref{(2.2.1)}) have the explicit azimuthal dependence of the schematic form	
\begin{equation}\label{(2.2.3)}{M^{A\dot B}} = \left( {\begin{array}{*{20}{c}}
  W&{X{e^{ - i\phi }}} \\ 
  {Y{e^{i\phi }}}&Z 
\end{array}} \right),\quad W,Z \in \Re ,\quad Y = {X^ * },\end{equation}
when the contribution of the term in $\xi $, which may be termed the primary spinor, is
\begin{eqnarray}\label{(2.2.4)}
  J & =& \left( {\begin{array}{*{20}{c}}
  {S{e^{i\chi }}}&{S{e^{ - i\chi }}} 
\end{array}} \right)\left( {\begin{array}{*{20}{c}}
  W&{X{e^{ - i\phi }}} \\ 
   {Y{e^{i\phi }}}&Z 
\end{array}} \right)\left( {\begin{array}{*{20}{c}}
  {S{e^{ - i\chi }}} \\ 
  {S{e^{i\chi }}} 
\end{array}} \right) \nonumber \\ 
 & = & {S^2}\left( {W + X{e^{2i\left( {\Theta  - mt} \right)}} + Y{e^{ - 2i\left( {\Theta  - mt} \right)}} + Z} \right) \\ 
& = & {S^2}\left( {W + 2{\rm Re} X\cos 2\left( {\Theta  - mt} \right) - 2{\rm Im} X\sin 2\left( {\Theta  - mt} \right) + Z} \right) ,\nonumber
\end{eqnarray}
clearly displaying an oscillatory dependence upon time.

Noting the way in which the azimuthal dependence of the IvdW1 ensures the absence of azimuthal dependence in the current, it is natural to incorporate into them a similar time dependence. This can be done by using the matrix
\begin{equation}\label{(2.2.5)}T := \left( {\begin{array}{*{20}{c}}
  {{e^{i\left( {\theta  - mt} \right)}}}&0 \\ 
  0&{{e^{ - i\left( {\theta  - mt} \right)}}} 
\end{array}} \right),\quad \theta  = {\theta ^ * } = \theta \left( {\rho ,z} \right),\end{equation}
to define modified forms that still satisfy the algebraic defining condition (\ref{(2.1.1)}),
\begin{eqnarray}\label{(2.2.6)}
 & \hat \sigma _A^{\mu \dot X} :=  T\sigma _A^{\mu \dot X}T;\quad \hat \sigma _A^{\mu \nu B} := \frac{1}{2}\left[ {\hat \sigma _A^{\mu \dot X}\hat \sigma _{\dot X}^{\nu B} - \hat \sigma _A^{\nu \dot X}\hat \sigma _{\dot X}^{\mu B}} \right] = T\sigma _A^{\mu \nu B}{T^ * } & \nonumber \\ 
   \Rightarrow & \hat \sigma _A^{\mu \dot X}\hat \sigma _{\dot X}^{\nu B} = T\left[ {\sigma _A^{\mu \dot X}\sigma _{\dot X}^{\nu B}} \right]{T^ * } = T\left[ { - \frac{1}{2}{g^{\mu \nu }}e_A^{ \bullet B} + \sigma _A^{\mu \nu B}} \right]{T^ * } =  - \frac{1}{2}{g^{\mu \nu }}e_A^{ \bullet B} + \hat \sigma _A^{\mu \nu B}. & 
\end{eqnarray}
At the same time as introducing time dependence, the opportunity has been taken in (\ref{(2.2.5)}) to include an arbitrary real phase function $ \theta $, and the redefined IvdW1 with upper spin indices are 
\begin{eqnarray}
  \hat \sigma _1^{A\dot B} = \frac{{ - 1}}{{\sqrt 2 }}\left( {\begin{array}{*{20}{c}}
  0&{{e^ - }} \\ 
  {{e^ + }}&0 
\end{array}} \right); & \hat \sigma _3^{A\dot B} = \frac{1}{{\sqrt 2 }}\left( {\begin{array}{*{20}{c}}
  { - 1}&0 \\ 
  0&1 
\end{array}} \right); & \quad {e^ \pm } := \exp  \pm 2i\left( {\Phi  + \theta } \right) \hfill  \nonumber \\
  \hat \sigma _2^{A\dot B} = \frac{{i\rho }}{{\sqrt 2 }}\left( {\begin{array}{*{20}{c}}
  0&{{e^ - }} \\ 
  { - {e^ + }}&0 
\end{array}} \right); & \hat \sigma _4^{A\dot B} = \frac{1}{{\sqrt 2 }}\left( {\begin{array}{*{20}{c}}
  1&0 \\ 
  0&1 
\end{array}} \right). & \hfill  \nonumber
\end{eqnarray}
These forms will no longer be covariantly constant as in (\ref{(2.1.4)}), and the reduction of the spinor identity in (\ref{(2.1.8)}) will therefore need to be modified. For the present however, it should be observed that (\ref{(2.2.3)}) and (\ref{(2.2.4)}) respectively become
\[{M^{A\dot B}} = \left( {\begin{array}{*{20}{c}}
  W&{X{e^ - }} \\ 
  {Y{e^ + }}&Z 
\end{array}} \right),\quad W,Z \in \Re ,\quad Y = {X^ * },\]
and
\begin{eqnarray}
  J & = &\left( {\begin{array}{*{20}{c}}
  {S{e^{i\chi }}}&{S{e^{ - i\chi }}} 
\end{array}} \right)\left( {\begin{array}{*{20}{c}}
  W&{X{e^ - }} \\ 
  {Y{e^ + }}&Z 
\end{array}} \right)\left( {\begin{array}{*{20}{c}}
  {S{e^{ - i\chi }}} \\ 
  {S{e^{i\chi }}} 
\end{array}} \right) \nonumber \\
 & = & {S^2}\left( {W + X{e^{2i\left( {\Theta  - \theta } \right)}} + Y{e^{ - 2i\left( {\Theta  - \theta } \right)}} + Z} \right) \nonumber \\ 
 & = & {S^2}\left( {W + 2{\rm Re} X\cos 2\left( {\Theta  - \theta } \right) - 2{\rm Im} X\sin 2\left( {\Theta  - \theta } \right) + Z} \right) .  \nonumber
\end{eqnarray} 
Time dependence has thus been eliminated, whilst the function $ \theta $ may be chosen as required to help ensure further appropriate properties for the current.
The time dependence incorporated here into the IvdW1 arises in a natural way from that proposed for the physical spinor, and as will appear below this will be true also in a similar way for the space dependence carried by $ \theta $. These modifications of the IvdW1 of course change the correspondence between global Poincaré transformations in spacetime and the local 2-spinor transformations associated with them. It is however by using the freedom present in (\ref{(2.1.1)}) in the form of the IvdW1 that consistency may be acheived in the observable flat spacetime context between the electromagnetic field of a spinor particle and the steady, axial flow of current responsible for that field. In this formulation the concept of intrinsic spin is not needed. 

Returning to the current, despite the simplifications arising from the potential having only non-vanishing azimuthal and time components, both independent of azimuth and time, reduction of (\ref{(2.2.1)}) is lengthy and complex, especially for the contribution of the secondary spinor, $ \zeta $. The results are 
\[ j_1^s = \frac{{\sqrt 2 {H^2}}}{{{m^2}}}\left( \begin{array}{c}
  2\left( {a2h3 - m\theta 1} \right) + \left[ \begin{array}{c}
  h{1^2} + {\left( {a4 + \Theta 3} \right)^2} \\ 
   - a{2^2} - 2{m^2} - h{3^2} - \Theta {1^2}  
\end{array}  \right]\cos 2\left( {\theta  - \Theta } \right) \\ 
   - 2\left[ {h3\left( {a4 + \Theta 3} \right) - h1\Theta 1} \right]\sin 2\left( {\theta  - \Theta } \right)  
\end{array}  \right),\]
\[ j_2^s = \frac{{\sqrt 2 {H^2}\rho }}{{{m^2}}}\left( \begin{array}{c}
  2\left( {a2\left( {a4 + \Theta 3} \right) - mh1} \right) - 2\left[ {h1\Theta 1 + h3\left( {a4 + \Theta 3} \right)} \right]\cos 2\left( {\theta  - \Theta } \right) \\ 
   - \left[ {a{2^2} - 2{m^2} - h{1^2} - h{3^2} + \Theta {1^2} + {{\left( {a4 + \Theta 3} \right)}^2}} \right]\sin 2\left( {\theta  - \Theta } \right)  
\end{array}  \right),\]
\[ j_3^s =  - \frac{{2\sqrt 2 {H^2}}}{{{m^2}}}\left( \begin{array}{c}
  a2h1 + m\left( {a4 + \Theta 3} \right) + \left[ {\theta 1\left( {a4 + \Theta 3} \right) - h1h3} \right]\cos 2\left( {\theta  - \Theta } \right) \\ 
   - \left[ {a2m + h3\Theta 1 + h1\left( {a4 + \Theta 3} \right)} \right]\sin 2\left( {\theta  - \Theta } \right) \\ 
\end{array}  \right),\]
\[ j_4^s = \frac{{\sqrt 2 {H^2}}}{{{m^2}}}\left( \begin{array}{c}
  a{2^2} + 2{m^2} + h{1^2} + h{3^2} + \Theta {1^2} + {\left( {a4 + \Theta 3} \right)^2} \\ 
   + 2\left[ {m\Theta 1 - a2h3} \right]\cos 2\left( {\theta  - \Theta } \right) - 2\left[ {mh1 + a2\left( {a4 + \theta 3} \right)} \right]\sin 2\left( {\theta  - \Theta } \right) \\ 
\end{array}  \right).\]
where the notation used here for computational convenience is
\[ \begin{array}{c}
  h1 := {\left( {\ln H} \right)_1} + 1/\left( {2\rho } \right) = {\left( {\ln \sqrt \rho  H} \right)_\rho };\;h3 := {\left( {\ln H} \right)_z}; \\ 
  \Theta 1 := {\Theta _\rho };\;\Theta 3 := {\Theta _z};\;\theta 1 := {\theta _\rho };\;\theta 3 := {\theta _z};\;a2 := q\frac{{{a_2}}}{\rho };a4 := q{a_4}.
\end{array} \]
Given that $ \theta $  may be freely chosen, it is possible to choose the difference of the two thetas to eliminate the $ \rho $-component of the primary spinor contribution to the current, proportional as it is to 
$\cos 2\left( {\theta -\Theta } \right)$. Choosing in particular that
\begin{equation}\label{(2.2.15)}\theta  - \Theta  = \frac{{\varepsilon \pi }}{4};\;\varepsilon  =  \pm 1,\end{equation}
the current components then become 
\begin{eqnarray}j_1^s & =& \frac{{2\sqrt 2 {H^2}}}{{{m^2}}}\left( {h3\left[ {a2 - \varepsilon \left( {a4 + \Theta 3} \right)} \right] + \Theta 1\left( {\varepsilon h1 - m} \right)} \right), \nonumber \\
j_2^s & =& \frac{{\sqrt 2 {H^2}\varepsilon \rho }}{{{m^2}}}\left( { - {{\left[ {a2 - \varepsilon \left( {a4 + \Theta 3} \right)} \right]}^2} + {m^2} + {{\left( {\varepsilon h1 - m} \right)}^2} + h{3^2} - \Theta {1^2}} \right), \nonumber\\
j_3^s & =& \frac{{2\sqrt 2 {H^2}\varepsilon }}{{{m^2}}}\left( {h3\Theta 1 - \left( {\varepsilon h1 - m} \right)\left[ {a2 - \varepsilon \left( {a4 + \Theta 3} \right)} \right]} \right), \nonumber\\
j_4^s & =& \frac{{\sqrt 2 {H^2}}}{{{m^2}}}\left( {{{\left[ {a2 - \varepsilon \left( {a4 + \Theta 3} \right)} \right]}^2} + {m^2} + {{\left( {\varepsilon h1 - m} \right)}^2} + h{3^2} + \Theta {1^2}} \right) . \nonumber\end{eqnarray}
Of the space components of the current, only the azimuthal can be non-vanishing, so it follows that, for an $H $ that is non-trivial, in the sense that $h_{3}$ is non-zero,
\begin{equation}\label{(2.2.20)}{\Theta _1} = 0 \wedge {\Theta _3} = \varepsilon a2 - a4 = q\left( {\varepsilon \frac{{{a_2}}}{\rho } - {a_4}} \right).\end{equation}
The spinor density then becomes
\begin{equation}\label{(2.2.21)}j_4^s = \frac{{\sqrt 2 {H^2}}}{{{m^2}}}\left( {{m^2} + {{\left( {\varepsilon h1 - m} \right)}^2} + h{3^2}} \right),\end{equation}
whilst there arises the relationship between spinor current components 
\begin{equation}\label{(2.2.22)}j_2^s = \varepsilon \rho j_4^s ,\end{equation}
a relation between four-vector components that will in general be referred to as {\em pseudoequality} and that here effectively expresses a uniform rotation of the spinor density ${j_4^s}$ about the ${z}$-axis, with the sense of rotation determined by the sign of  ${\varepsilon}$.

\subsection{Equality of Currents in Poincaré Spacetime}\label{sec:poincarecurrents}

Whilst the treatment of the spinor current in the previous Section was complete and consistent, it will now be shown that the conditions on which it was based lead to an electromagnetic current that in a Poincaré spacetime cannot be identified with anything other than a trivial spinor matter current. Consider in particular (\ref{(2.2.20)}), which imposes near-pseudoequality on the potentials, 
\begin{equation}\label{(2.3.1)}\frac{{{a_2}}}{\rho } = \varepsilon \left( {{a_4} + \frac{{{\Theta _3}}}{q}} \right),\end{equation}
where ${\Theta _3}$ is an arbitrary function of $z$. Using this now with electromagnetic current components from (\ref{(2.1.11)}) and the pseudoequality in (\ref{(2.2.22)}), if electromagnetic and spinor matter currents are essentially identified, then it follows that

\begin{equation}\label{(2.3.2)}\left\{ {{{\vec \nabla }^2} - \frac{1}{{{\rho ^2}}}} \right\}\frac{{{a_2}}}{\rho } = \varepsilon {\vec \nabla ^2}{a_4} \Leftrightarrow \left\{ {{{\vec \nabla }^2} - \frac{1}{{{\rho ^2}}}} \right\}\left( {{a_4} + \frac{{{\Theta _3}}}{q}} \right) = {\vec \nabla ^2}{a_4} \Leftrightarrow {a_4} = {\rho ^2}d_z^2\frac{{{\Theta _3}}}{q} - \frac{{{\Theta _3}}}{q}.\end{equation}
The only physically consistent solution of this condition is that in which matter and electromagnetic currents both vanish. The Poincaré spacetime is therefore empty, and the geometrisation project reported here must be deemed to have failed. 

In an effort to recover the project, consider again the second of the three equivalent conditions in (\ref{(2.3.2)}) which, with ${\Theta _3}$ taken to vanish for simplicity, requires that 
\begin{equation}\label{(2.3.3)}\left\{ {{{\vec \nabla }^2} - \frac{1}{{{\rho ^2}}}} \right\}{a_4} = {\vec \nabla ^2}{a_4}.\end{equation}
In regions distant from the $z$-axis the $\frac{1}{{{\rho ^2}}}$ here (almost) vanishes, when the difficulty disappears and (\ref{(2.3.3)}) essentially becomes an identity. This suggests that perhaps the differential operators here, especially that with the $\frac{1}{{{\rho ^2}}}$ term that is singular along the $z$-axis, may be giving an inadequate account of the derivatives that would appear were the underlying spacetime to exhibit torsion. An earlier geometrisation investigation within conformally Minkowskian QED \cite{leigh2} already demonstrated a need for torsion, whilst treatments using General Relativity \cite{hehl,leigh1} also suggest a natural association between spin and torsion.  In line with these considerations, Section~\ref{sec:cochlean} introduces a specific torsion, replacing Poincaré spacetime by what will there be called {\em Cochlean spacetime}.\footnote{The author favours the suggestive coinage ‘Cochlean’ over more prosaic appellations such as, say, ‘axial’, from the subgroup of space transformations under which a Cochlean spacetime is invariant.}  In regions distant from the z-axis where $\frac{1}{{{\rho ^2}}}$ (almost) vanishes, Cochlean spacetime must, of course, be indistinguishable from its Poincaré precursor.

Before introducing torsion however, it is worth examining more closely the status of the Poincaré pseudoequality conditions, (\ref{(2.2.22)}) and (\ref{(2.3.1)}), and what it means to insist on their each being satisfied.
To begin with, the pseudoequality of potential components appearing in (\ref{(2.3.1)}) [with ${\Theta _3}$ vanishing] not only decouples electromagnetism in the spinor identity, with terms in ${a_2}$ and ${a_4}$ cancelling, it obviously precludes any treatment based upon lower multipoles, such as a Coulomb monopole with, say, a magnetic dipole. Leaving aside (unbounded) multipoles then, consider an example of an everywhere-bounded functional form such as
\begin{equation}\label{(2.3.4)}{a_4} = \varepsilon \frac{{{a_2}}}{\rho } =  - \frac{{q\rho }}{{{r^2}}}\exp \left( { - \frac{\mu }{{qr}}} \right),\;\mu q > 0,\quad r := \sqrt {{\rho ^2} + {z^2}}. \end{equation}
Although pseudoequality of potentials clearly imposes a close relationship between the electric and magnetic fields of a putative static fermion, for asymptotically large $\rho$ the components in this example have the quasi-multipole forms 
\[ {a_2} =  - \varepsilon q\left( {1 - \frac{{\mu {\rho ^2}}}{{q{r^3}}} + O\left( {\frac{{{z^2}}}{{{\rho ^2}}}} \right)} \right),\quad {a_4} = \left( {\frac{{ - q}}{r} + O\left( {\frac{z}{\rho }} \right)} \right)\left( {1 - \frac{\mu }{{qr}} + O\left( {\frac{1}{{{\rho ^2}}}} \right)} \right),\]
where the independent parameters $q$ and $\mu$ appearing in (\ref{(2.3.4)}) determine the total charge and magnetic moment respectively.
This simple example shows that there is no essential difficulty in the idea of electrostatic and magnetostatic fields being generated by pseudoequal potential components. 

The real difficulty arises when one seeks to re-couple electromagnetism by insisting that the spinor current with pseudoequal components (\ref{(2.2.22)}) generate an electromagnetic field that has pseudoequal potentials (\ref{(2.3.1)}). Both the current given by the electromagnetic potentials and that generated by the spinor field are Poincaré conserved, and it is natural to wish to identify them, an essential and automatic feature of QED, but one that here leads to (\ref{(2.3.2)}), showing this natural identification to be impossible in a Poincaré spacetime. An alternative geometry is required.

 \section{The Spinor Identity and its Solutions in Cochlean Spacetime}\label{sec:cochlean}

 \subsection{The Cochlean Connection}

The most general tensor connection under which a metric tensor $g_{\mu \nu}$ is constant is the sum of the torsion-free Levi-Cività form and the contortion tensor
\[{\gamma _{\lambda \mu \nu }} = {\tilde \gamma _{\lambda \mu \nu }} + {c_{\lambda \mu \nu }},\quad {\tilde \gamma _{\lambda \mu \nu }} = \frac{1}{2}\left( {{g_{\lambda \mu \nu }} + {g_{\lambda \nu \mu }} - {g_{\mu \nu \lambda }}} \right),\]
where contortion may be defined in terms of torsion by\footnote{These standard results appear in Popławski\cite{poplawski}, for example.}
\[ {c_{\lambda \mu \nu }} := \frac{1}{2}\left( {{t_{\mu \nu \lambda }} - {t_{\lambda \nu \mu }} + {t_{\nu \mu \lambda }}} \right),\quad {t_{\lambda \mu \nu }} := {\gamma _{\lambda \mu \nu }} - {\gamma _{\lambda \nu \mu }}.\]
The standard cylinder metric 
${g_{\mu \nu }} = \left( { - 1, - {\rho ^2}, - 1,1} \right)$ gives the Levi-Cività connections used in the previous Section, namely
\[ {\tilde \gamma _{\lambda \nu \mu }} = \rho \left[ {{\delta _{\nu 2}}{\delta _{\lambda \mu 12}} - {\delta _{\nu 1}}{\delta _{\lambda 2}}{\delta _{\mu 2}}} \right] = \rho \left[ {{\delta _{\mu 2}}{\delta _{\lambda \nu 12}} - {\delta _{\mu 1}}{\delta _{\lambda 2}}{\delta _{\nu 2}}} \right],\]
where the earlier definition that ${\delta _{\alpha \beta \theta \phi }} := {\delta _{\alpha \theta }}{\delta _{\beta \phi }} - {\delta _{\alpha \phi }}{\delta _{\beta \theta }}$ has been used.

An ad-hoc form for the contortion will now be hypothesised. To begin with, if the spin axis of the fermion is to play a central role, then contortion should be chosen so as not to influence derivative forms with respect to the third and fourth coordinates $z$ and $t$, translations with respect to which should leave spacetime invariant. At the same time, physical consistency requires that the contortion tensor vanish at large distances from the fermion spin axis, in what will be termed the {\em asymptotic region} where Poincaré and putative Cochlean spacetimes locally coincide. These requirements, together with simple dimensional constraints and skew symmetry on the leading indices, suggest the choice 
\begin{equation}\label{(3.1.4)}{c_{\lambda \nu \mu }} =  - \rho {\kern 1pt} {\delta _{\mu 2}}{\delta _{\lambda \nu 12}}.\end{equation}
A dimensionless constant factor could be included here, but that would produce a more complicated Cochlean connection, whilst the form (\ref{(3.1.4)}) gives the strikingly simple
\begin{equation}\label{(3.1.5)}\breve{\gamma } _{ \bullet \nu \mu }^\lambda  := \tilde \gamma _{ \bullet \nu \mu }^\lambda  + c_{ \bullet \nu \mu }^\lambda  = \frac{1}{\rho }\delta _2^\lambda {\delta _{\nu 2}}{\delta _{\mu 1}};\quad \breve{t} _{ \bullet \mu \nu }^\lambda  := \breve{\gamma } _{ \bullet \mu \nu }^\lambda  - \breve{\gamma } _{ \bullet \nu \mu }^\lambda  = \frac{1}{\rho }\delta _2^\lambda {\delta _{\mu \nu 21}}.\end{equation}
From now on tilde and breve diacritical marks will be used as required to distinguish respectively between Poincaré and Cochlean tensor connections and objects related to them. 
When torsion is present in a Riemann space it is necessary\footnote{For the torsion subtraction here see Hehl\cite{hehl} or Popławski\cite{poplawski}.} to define the curvature tensor using the commutator of derivatives with a torsion subtraction, which in the Cochlean case gives
\begin{eqnarray}
  2 {\breve{ d }}_{[\alpha }\breve{d}_{\beta ]}v_{\eta } - \breve{t} _{ \bullet \alpha \beta }^{\sigma} \breve{d}_ {\sigma }v_{\eta }& =& 2 \breve{d} _{[\alpha } \breve{d} _{\beta ]} v_{\eta }  - \frac{1}{\rho }\delta _2^\sigma {\delta _{\alpha \beta 21}}\left( {{d_\sigma }{v_\eta } -  \breve{\gamma } _{ \bullet \eta \sigma }^\varepsilon {v_\varepsilon }} \right)  \nonumber\\  &  =&  2 \breve {d} _{[\alpha }}{\breve {d} _{\beta ]}{v_\eta } - \frac{1}{\rho }\delta _2^\sigma {\delta _{\alpha \beta 21}}\left( {{d_\sigma }{v_\eta } - \frac{1}{\rho }\delta _2^\varepsilon {\delta _{\eta 2}}{\delta _{\sigma 1}}{v_\varepsilon }} \right)  \nonumber \\ 
  & =& 2{{\breve{d} }_{[\alpha }}{{\breve{d} }_{\beta ]}}{v_\eta } - \frac{{{d_2}{v_\eta }}}{\rho }{\delta _{\alpha \beta 21}} = 0, \nonumber
\end{eqnarray}
and the Cochlean curvature tensor, like its Poincaré counterpart, therefore vanishes. In the standard terminology, both Poincaré and Cochlean spacetimes are {\em flat}.

As to the status of the spinor identity, it is clear that the Poincaré spinor formulation given in  (\ref{(2.1.1)}) – (\ref{(2.1.3)}) is still valid in Cochlean space, and the steps that culminate in the Poincaré form (\ref{(2.1.10)}) require modification only in that the covariant derivative $D$ must now be based upon the Cochlean connection. 
The treatment of the spinor current is also essentially unchanged, since it is only first-order derivatives [of spinors] that appear in it. It should be noted that the modification of the IvdW1 (\ref{(2.2.6)}) that appears in the Poincaré current may be used in the Cochlean case for the same reason as that for which it was originally introduced, namely so as to accommodate a static electromagnetic field in the spinor identity. Since the spinor current treatment remains unchanged, conditions (\ref{(2.2.15)}) and (\ref{(2.2.20)}) that ensure correct physical forms for the current components are still valid.

\subsection{Cochlean Electromagnetism}

In analogy with the Poincaré case, a Cochlean electromagnetic field may be defined by [omitting the breve on the Cochlean potential]
\begin{equation}\label{(3.2.1)}\breve{f} _{\mu \nu } := \breve{d} _{\mu }{a_\nu } - \breve{d} _{\nu } a_{\mu } = {d_\mu }{a_\nu } - {d_\nu }{a_\mu } + \breve{\tau } _{ \bullet \mu \nu }^\sigma {a_\sigma } = {d_\mu }{a_\nu } - {d_\nu }{a_\mu } - \frac{{{a_2}}}{\rho }\,{\delta _{\mu \nu 12}},\end{equation}
resulting, under current assumptions about the potential, in the forms
\begin{equation}\label{(3.2.2)}\breve{f} _{\mu \nu } = \left( {\begin{array}{*{20}{c}}
  0&{\rho {d_1}\frac{{{a_2}}}{\rho }}&0&{{d_1}{a_4}} \\ 
  { - \rho {d_1}\frac{{{a_2}}}{\rho }}&0&{ - {d_3}{a_2}}&0 \\ 
  0&{{d_3}{a_2}}&0&{{d_3}{a_4}} \\ 
  { - {d_1}{a_4}}&0&{ - {d_3}{a_4}}&0 
\end{array}} \right) = \left( {\begin{array}{*{20}{c}}
  0&{\rho {B_3}}&0&{ - {E_1}} \\ 
  { - \rho {B_3}}&0&{\rho {B_1}}&0 \\ 
  0&{ - \rho {B_1}}&0&{ - {E_3}} \\ 
  {{E_1}}&0&{{E_3}}&0 
\end{array}} \right).\end{equation}
The electromagnetic current, conventionally the divergence of this field, is now
\[ \breve{j} _\nu ^e = {\delta _{\nu 4}}{g^{\mu \sigma }}{d_\sigma }{d_\mu }{a_\nu } + {\delta _{\nu 2}}\rho {g^{\mu \sigma }}{d_\sigma }{d_\mu }\left( {\frac{{{a_\nu }}}{\rho }} \right),\]
and is therefore given by the harmonic operator ${d^2} := d_1^2 + d_3^2$  acting on the respective potential components. It follows that if these components satisfy the pseudoequality relation (which requires the choice $\Theta_{3}=0$)
\[{\breve{a} _2} = \varepsilon \rho {\breve{a} _4},\]
then the Cochlean electromagnetic current equally satisfies
\begin{equation}\label{(3.2.5)}{\breve{\jmath} _2} = \varepsilon \rho {\breve{\jmath} _4}.\end{equation}
It is apparent from (\ref{(3.1.5)}) and the lack of time and azimuthal dependence of the current components that this current is Cochlean conserved.
The structure reviewed here therefore offers a simple resolution of the problem of the lack of conformality between the Poincaré electromagnetic current and the Poincaré spinor current. 

To complete the Cochlean treatment of electromagnetism, consider the energy-momentum density tensor that may be defined, following the Poincaré model, by 
\[ \breve{t} _{\mu \nu }^e := \breve{f} _\mu ^{ \bullet \alpha }{\breve{f} _{\alpha \nu }} + \frac{1}{4}{g_{\mu \nu }}{\breve{f} ^{\alpha \beta }}{\breve{f} _{\alpha \beta }},\]
in which pseudoequality of the potentials causes the trace subtraction to vanish:
\begin{eqnarray}
  {g^{\alpha \gamma }}{g^{\beta \delta }}{{\breve{f} }_{\alpha \beta }}{{\breve{f} }_{\gamma \delta }}& =& 2\left[ {\frac{1}{{{\rho ^2}}}{{\breve{f} }_{12}}{{\breve{f} }_{12}} - {{\breve{f} }_{14}}{{\breve{f} }_{14}} + \frac{1}{{{\rho ^2}}}{{\breve{f} }_{32}}{{\breve{f} }_{32}} - {{\breve{f} }_{34}}{{\breve{f} }_{34}}} \right] \nonumber \\ 
  & =& 2\left[ {{{\left( {{d_1}\frac{{{a_2}}}{\rho }} \right)}^2} - {{\left( {{d_1}{a_4}} \right)}^2} + {{\left( {{d_3}\frac{{{a_2}}}{\rho }} \right)}^2} - {{\left( {{d_3}{a_4}} \right)}^2}} \right] = 0.  \nonumber\end{eqnarray}
It can now be seen that the electromagnetic energy-momentum density has non-zero Cochlean components only for indices that are either 2 or 4, and that the forms are
\begin{equation}\label{(3.2.8)}\breve{t} _{44}^e = {g^{\alpha \beta }}{\breve{f} _{4\alpha }}{\breve{f} _{\beta 4}} = {\left( {{d_1}{a_4}} \right)^2} + {\left( {{d_3}{a_4}} \right)^2},\end{equation}
\begin{equation}\label{(3.2.9)}\breve{t} _{24}^e = \breve{t} _{42}^e = {g^{\alpha \beta }}{\breve{f} _{2\alpha }}{\breve{f} _{\beta 4}} = \left( {\rho {d_1}\frac{{{a_2}}}{\rho }} \right)\left( {{d_1}{a_4}} \right) + \left( {\rho {d_3}\frac{{{a_2}}}{\rho }} \right)\left( {{d_3}{a_4}} \right) = \varepsilon \rho {\breve{t} _{44}},\end{equation}
\begin{equation}\label{(3.2.10)}\breve{t} _{22}^e = {g^{\alpha \beta }}{\breve{f} _{2\alpha }}{\breve{f} _{\beta 2}} = {\left( {\rho {d_1}\frac{{{a_2}}}{\rho }} \right)^2} + {\left( {\rho {d_3}\frac{{{a_2}}}{\rho }} \right)^2} = {\rho ^2}{\breve{t} _{44}}.\end{equation}
This tensor is also conserved, 
\[ {g^{\sigma \mu }}{\breve{d} _\sigma }\breve{t} _{\mu \nu }^e = {\breve{d} _4}\breve{t} _{4\nu }^e - \frac{1}{{{\rho ^2}}}{\breve{d} _2}\breve{t} _{2\nu }^e = 0.\]

\subsection{Reducing and Solving the Cochlean Identity}\label{sec:cochleanidentity}

The basic Poincaré spinor formulation given in  (\ref{(2.1.1)})–(\ref{(2.1.4)}) remains unchanged in a Cochlean space, as does the treatment of the spinor current, since only first-order derivatives of spinors appear in it. As noted earlier, since treatment of the spinor current is unchanged, conditions (\ref{(2.2.15)}) and (\ref{(2.2.20)}) ensuring the correct physical forms for the current components remain valid.
As to the status of the spinor identity, the steps that culminate in the Poincaré form (\ref{(2.1.10)}) require modification in the Cochlean case only in that the covariant derivative $D$ must be based upon the Cochlean connection, rather than the Poincaré one. The second-order derivative that appears in the identity can then be expanded as follows:
\begin{eqnarray}
  D_A^{ \bullet \dot X}D_{\dot X}^{ \bullet B}{\xi _B}& =& \hat \sigma _A^{\mu \dot X}{D_\mu }\left( {\hat \sigma _{\dot X}^{\nu B}{D_\nu }{\xi _B}} \right) = \hat \sigma _A^{\mu \dot X}\left( {{D_\mu }\hat \sigma _{\dot X}^{\nu B}} \right){D_\nu }{\xi _B} + \hat \sigma _A^{\mu \dot X}\hat \sigma _{\dot X}^{\nu B}\left( {{D_\mu }{D_\nu }{\xi _B}} \right) \nonumber\\ 
 &  =& \hat \sigma _A^{\mu \dot X}\left( {{{\breve{d} }_\mu }\hat \sigma _{\dot X}^{\nu B} - 2iq{a_\mu }\hat \sigma _{\dot X}^{\nu B}} \right){D_\nu }{\xi _B} + \left( {\hat \sigma _A^{\mu \nu B} - \frac{1}{2}{g^{\mu \nu }}e_A^{ \bullet B}} \right){D_\mu }{D_\nu }{\xi _B} \nonumber \\ 
   & =& \hat \sigma _A^{\mu \dot X}\left( {{{\breve{d} }_\mu }\hat \sigma _{\dot X}^{\nu B}} \right){D_\nu }{\xi _B} + \left( {\hat \sigma _A^{\mu \nu B} - \frac{1}{2}{g^{\mu \nu }}e_A^{ \bullet B}} \right)\left( {{D_\mu } - 2iq{a_\mu }} \right){D_\nu }{\xi _B} \nonumber \\ 
  & =& \left( {\hat \sigma _A^{\mu \nu B} - \frac{1}{2}{g^{\mu \nu }}e_A^{ \bullet B}} \right)\left( {{{\breve{d} }_\mu } - iq{a_\mu }} \right)\left( {{d_\nu } + iq{a_\nu }} \right){\xi _B} + \hat \sigma _A^{\mu \dot X}\left( {{{\breve{d} }_\mu }\hat \sigma _{\dot X}^{\nu B}} \right){D_\nu }{\xi _B}.  \nonumber \end{eqnarray}
The Cochlean derivative appearing here does not include spinor connections, and the expansion allows the Cochlean spinor identity to be written in a calculationally convenient form as an expression for the coupling term 
\begin{equation}\label{(3.3.2)}iq\breve{f} _A^B{\xi _B} = {F_1} - {F_2};\quad \left\langle {\breve{f} _A^B := {{\breve{f} }_{\mu \nu }}\hat \sigma _A^{\mu \nu B}} \right\rangle,\end{equation}
with
\begin{eqnarray}
  {F_1}& := &\left[ {{g^{\mu \nu }}\left( {{{\breve{d} }_\mu } - iq{a_\mu }} \right)\left( {{d_\nu } + iq{a_\nu }} \right) + {m^2}} \right]{\xi _A}, \hfill \nonumber \\
  {F_2}& := &2\hat \sigma _A^{\mu \dot X}\left( {{{\breve{d} }_\mu }\hat \sigma _{\dot X}^{\nu B}} \right){D_\nu }{\xi _B} - \frac{2}{\rho }\hat \sigma _A^{12B}{d_2}{\xi _B} - 4iq\hat \sigma _A^{\mu \nu B}{a_\mu }{d_\nu }{\xi _B}. \hfill  
\nonumber \end{eqnarray} 
The LHS of (\ref{(3.3.2)}) is a linear expression in the first derivatives of the components of the electromagnetic potential whilst, noting that ${g^{\mu \nu }}{d_\mu }{a_\nu } = 0$  in $F_{1}$, the RHS has no terms in such derivatives; solving the identity requires that these forms be reconciled. 
In constructing the spinor form of the electromagnetic field for the LHS of (\ref{(3.3.2)}), the redefined IvdW2 that are needed can be calculated directly from products of the transformed (\ref{(2.1.4)}), which are
\begin{eqnarray}
  \hat \sigma _A^{1\dot B}& =& \frac{1}{{\sqrt 2 }}\left( {\begin{array}{*{20}{c}}
  { - {e ^ + }}&0 \\ 
  0&{{e ^ - }} 
\end{array}} \right);\;\hat \sigma _A^{3\dot B} = \frac{1}{{\sqrt 2 }}\left( {\begin{array}{*{20}{c}}
  0&1 \\ 
  1&0 
\end{array}} \right);\quad { e^ \pm } = \exp  \pm 2i\left( {\Phi  + \theta } \right) \hfill \nonumber \\
  \hat \sigma _A^{2\dot B}& =& \frac{{ - i}}{{\sqrt 2 \rho }}\left( {\begin{array}{*{20}{c}}
  {{e ^ + }}&0 \\ 
  0&{{e ^ - }} 
\end{array}} \right);\;\hat \sigma _A^{4\dot B} = \frac{1}{{\sqrt 2 }}\left( {\begin{array}{*{20}{c}}
  0&{ - 1} \\ 
  1&0 
\end{array}} \right), \hfill \nonumber
\end{eqnarray} 
and these give
\begin{eqnarray}
  \hat \sigma _A^{12B}& =& \frac{i}{{2\rho }}\left( {\begin{array}{*{20}{c}}
  { - 1}&0 \\ 
  0&1 
\end{array}} \right);\;\hat \sigma _A^{13B} = \frac{1}{2}\left( {\begin{array}{*{20}{c}}
  0&{ - {e^ + }} \\ 
  {{e^ - }}&0 
\end{array}} \right);\;\hat \sigma _A^{14B} = \frac{1}{2}\left( {\begin{array}{*{20}{c}}
  0&{{e^ + }} \\ 
  {{e^ - }}&0 
\end{array}} \right); \hfill \nonumber \\
  \hat \sigma _A^{23B}& =& \frac{{ - i}}{{2\rho }}\left( {\begin{array}{*{20}{c}}
  0&{{e^ + }} \\ 
  {{e^ - }}&0 
\end{array}} \right);\;\hat \sigma _A^{24B} = \frac{i}{{2\rho }}\left( {\begin{array}{*{20}{c}}
  0&{{e^ + }} \\ 
  { - {e^ - }}&0 
\end{array}} \right);\;\;\hat \sigma _A^{34B} = \frac{1}{2}\left( {\begin{array}{*{20}{c}}
  1&0 \\ 
  0&{ - 1} 
\end{array}} \right). \hfill  \nonumber
\end{eqnarray} 

Explicit calculation using the definitions from (\ref{(3.2.1)}) and (\ref{(3.2.2)}) now gives the spinor field to be
\begin{eqnarray}
  \breve{f} _A^B& =& 2{{\breve{f} }_{12}}\hat \sigma _A^{12B} + 2{{\breve{f} }_{14}}\hat \sigma _A^{14B} + 2{{\breve{f} }_{23}}\hat \sigma _A^{23B} + 2{{\breve{f} }_{34}}\hat \sigma _A^{34B} \nonumber \\ 
  & =& i{d_1}\frac{{{a_2}}}{\rho }\left( {\begin{array}{*{20}{c}}
  { - 1}&0 \\ 
  0&1 
\end{array}} \right) + {d_1}{a_4}\left( {\begin{array}{*{20}{c}}
  0&{{e^ + }} \\ 
  {{e^ - }}&0 
\end{array}} \right) + \frac{i}{\rho }{d_3}{a_2}\left( {\begin{array}{*{20}{c}}
  0&{{e^ + }} \\ 
  {{e^ - }}&0 
\end{array}} \right) + {d_3}{a_4}\left( {\begin{array}{*{20}{c}}
  1&0 \\ 
  0&{ - 1} 
\end{array}} \right) .  \nonumber
\end{eqnarray}
Before taking the product of this with the physical spinor it should be noted that, with 
\[{e^ \pm } = \exp  \pm 2i\left( {\Phi  + \theta } \right) = \exp  \pm 2i\left( {\Phi  + \Theta  + \varepsilon \pi /4} \right) =  \pm i\varepsilon \exp  \pm 2i\chi ,\]
it becomes clear that the expression for the coupling term can involve only the traceless matrix $\hat 1 := \left( {\begin{array}{*{20}{c}} 1&0\\ 0&{ - 1}\end{array}} \right)$:
\begin{equation}\label{(3.3.8)}\breve {f} _A^B{\xi _B} =  - i\left( {{d_1}\frac{{{a_2}}}{\rho }} \right)\hat 1{\xi _B} + i\varepsilon \left( {{d_1}{a_4}} \right)\hat 1{\xi _B} - \frac{\varepsilon }{\rho }\left( {{d_3}{a_2}} \right)\hat 1{\xi _B} + \left( {{d_3}{a_4}} \right)\hat 1{\xi _B} =  - \frac{{{\Theta _{33}}}}{q}\hat 1{\xi _B}.\end{equation}

Turning now to the RHS of (\ref{(3.3.2)}), straightforward technical simplification gives 
\[ {F_1} = \left[ \begin{array}{c}
  \Theta _3^2 - \frac{{{d^2}H}}{H} \\ 
   + \frac{1}{{4{\rho ^2}}} + {q^2}\left( {a_4^2 - \frac{{a_2^2}}{{{\rho ^2}}}} \right) \\ 
\end{array}  \right]{\xi _A} - i\left[ {{\Theta _{33}} + 2{\Theta _3}{h_3}} \right]\hat 1{\xi _A},\]
\[{F_2} = \left[ \begin{array}{c}
  \frac{{{H_1}}}{H}\left( {\frac{1}{\rho } - 2\varepsilon m} \right) - \varepsilon \frac{m}{\rho } \\ 
   + 2q\varepsilon {\kern 1pt} {\Theta _3}\frac{{{a_2}}}{\rho } + 2{q^2}\left( {a_4^2 - \frac{{a_2^2}}{{{\rho ^2}}}} \right) \\ 
\end{array}  \right]{\xi _A} - 2i{\kern 1pt} {\Theta _3}{h_3}\hat 1{\xi _B}.\]
The contribution here involving the traceless matrix is
\[{\left. {\left( {{F_1} - {F_2}} \right)} \right|_{traceless}} =  - i{\Theta _{33}}\hat 1 ,\]
consonant with (\ref{(3.3.8)}) so that, provided the contribution to the identity matrix factor from $F_{1}-F_{2}$ vanishes, the spinor identity will be satisfied with any form for $\Theta _{3}$. It is in taking advantage of this observation that the simple choice of vanishing $\Theta _{3}$ has been made above; as already noted, this removes direct coupling between the spinor field and its electromagnetic potential through the spinor identity. Indeed, because of the form of the IvdW1 with spacetime indices 2 and 4, as has already been pointed out, electromagnetism decouples when a single spinor derivative operates on $\xi _{A}$, with pseudoequality causing the terms in $a_{2}$ and  $a_{4}$ to cancel. 
With the vanishing of $\Theta _{3}$, there then follows for the identity matrix contribution
\[ \mathop {{{\breve{F} }_1} - {{\breve{F} }_2}}\limits_{identity}  =  - \left[ {\frac{{{{\vec \nabla }^2}H}}{H} - \frac{1}{{4{\rho ^2}}} - 2\varepsilon m{h_1}} \right]{\xi _A},\quad {h_1} = {d_\rho }\ln \left( {\sqrt \rho  H/\left| m \right|} \right).\]
The second-order differential equation that results from setting this to zero is then a necessary and sufficient condition for the physical spinor itself to be a solution of the Cochlean spinor identity. Exact forms for solutions can be obtained by first making a change of variables [which will require $\varepsilon m$ to be negative],
\[ T\left( {x,y} \right) := \sqrt \rho  H\left( {\rho ,z} \right),\quad x =  - 2\varepsilon m\rho ,\;y =  - 2\varepsilon mz,\]
when there follows 
\begin{equation}\label{(3.3.14)}\left( {d_x^2 + {d_x} + d_y^2} \right)T = 0.\end{equation}
A simple approach to solving this is to separate with $T(x,y) = X(x)Y(y)$, when standard procedures lead to two possible forms of separated solution, to wit
\[ T = {T_0}\exp \frac{{ - 1}}{2}\left( {\left( {1 - \cos \delta } \right)x + \sin \delta y} \right),\quad \delta  \in \left] {0,\;\pi } \right[,\]
\[ T = {T_0}\left( {1 + \eta x} \right)\exp \left( { - \frac{{x + y}}{2}} \right),\]
and these coincide when the parameters $\delta $ and $\eta $ take the values $\pi/2 $ and 0 respectively. 

In general, the exponents in these two forms of solution differ significantly, being respectively in coordinate variables 
\begin{equation}\label{(3.3.23)} - \left| m \right|\left( {\left( {1 - \cos \delta } \right)\rho  + \sin \delta \left| z \right|} \right){\rm  and  } - \left| m \right|\left( {\rho  + \left| z \right|} \right)\quad \left\langle {\left| m \right| =  - \varepsilon m} \right\rangle, \end{equation}
and the two corresponding types of solution will be called in this paper ‘moderated’ and ‘unmoderated’ respectively, in reference to moderation of the exponential decay in $\rho $ and $z $ by trigonometric functions of $\delta $ in solutions of the first type. 
Noting the nature of the spatial decay in the asymptotic region suggests that a moderated solution is associated with a fermion of which the mass is given by
\begin{equation}\label{(3.3.24)}{m_\ell} = \left( {1 - \cos \delta } \right)\left| m \right|.\end{equation}

Casting around now for a suitable mass scale, and bearing in mind the desirability of interpreting the current Cochlean approach as a simplification to flat spaces of a treatment originally formulated within General Relativity, a natural choice is that the modulus of $m $ should be taken to be $M_P $, the Planck mass that appears in the Einstein equation in the form
\[ {T_{44}} = \frac{{{E_{44}}}}{{8\pi {\kern 1pt} G}} = {\kern 1pt} \frac{{M_P^2}}{{8\pi }}{\kern 1pt} {E_{44}}.\]
Such a choice however brings in the well-known problem of scale expressed by
\[ {M_P} = \sqrt {\frac{{\hbar c}}{G}}  \approx 2.2 \times {10^{ - 8}}{\rm kg},{M_e} \approx 9.1 \times {10^{ - 31}}{\rm kg} \sim {10^{ - 23}}{M_P},\]
where $M_e $ is the electron mass. There is some hope of coping with this problem by choosing values of $\delta $  extremely close to zero in the first of (\ref{(3.3.23)}), enabling moderated solutions to describe stable fermions using the same spinor identity, with particles of differing masses associated with exponential decays moderated by differing values of $\delta $. Solutions with unmoderated exponentials may then be expected to define states with very high mass. Since such particles have not been observed, it may be assumed that unmoderated solutions are electromagnetically inert; although potentially contributing to dark matter, such states are not of interest for current purposes and will not therefore be treated further here. The mass parameter will be discussed in a subsequent paper that employs the framework of General Relativity, whilst the form $m $ will be retained here on the provisional understanding that its modulus is assumed to be $M_P $, although all Cochlean results remain valid whatever non-zero value $m $ has.

\subsection{Normalising the Spinor Density} \label{sec:normspindensity}

For moderated solutions, the spinor modulus has the simple separated form
\begin{equation}\label{(3.4.1)}H = {H_0}\frac{{{{\left| m \right|}^{3/2}}}}{{\sqrt x }}{e^{ - r}},\quad r := \frac{1}{2}\left( {x\left( {1 - \cos \delta } \right) + \left| y \right|\sin \delta } \right),\quad \delta  \in \left] {0,\;\pi } \right[,\end{equation}
where $H_0 $ is a dimensionless normalisation constant that will now be determined.
The spinor density is given by (\ref{(2.2.1)}) as the sum of two contributions; that from the primary spinor, given by 
\begin{equation}\label{(3.4.2)}{\left. {j_4^s} \right|_{primary}} = \sigma _4^{A\dot B}{\xi _A}{\xi _{\dot B}} = \sqrt 2 {H^2},\end{equation}
with (\ref{(2.2.21)}) then implying that the secondary contribution is
\begin{equation}\label{(3.4.3)}{\left. {j_4^s} \right|_{secondary}} = \sigma _4^{A\dot B}{\zeta _A}{\zeta _{\dot B}} = \frac{{{{\left( {\varepsilon h1 - m} \right)}^2} + h{3^2}}}{{{m^2}}}\sqrt 2 {H^2}.\end{equation}
The detailed form of the secondary spinor given in (\ref{(2.1.5)}) can be written
\begin{eqnarray}{\zeta _A}& =& \frac{H}{m}\left( {\begin{array}{*{20}{c}}
{\left( \begin{array}{c}
h1 - i\Theta 1\\
 + a2
\end{array} \right){e^ + }}&\begin{array}{c}
 - h3 - i\Theta 3\\
 - ia4 - im
\end{array}\\
\begin{array}{c}
 - h3 + i\Theta 3\\
 + ia4 - im
\end{array}&{\left( \begin{array}{c}
 - h1 - i\Theta 1\\
 + a2
\end{array} \right){e^ - }}
\end{array}} \right)\left( {\begin{array}{*{20}{c}}
{{e^{ - i\chi }}}\\
{{e^{i\chi }}}
\end{array}} \right) \nonumber \\& =& \frac{{i\varepsilon h1 - im - h3}}{m}{\xi _A}\nonumber \end{eqnarray}
when the origin of (\ref{(3.4.3)}) becomes evident. Specifically for simple, moderated solutions of form (\ref{(3.4.1)}), it follows that
\[ h1 := {d_1}\ln \left( {\sqrt \rho  H} \right) = \varepsilon m\left( {1 - \cos \delta } \right),\quad h3 := {d_3}\ln \left( {\sqrt \rho  H} \right) = \varepsilon m\sin \delta ,\]
so that
\[ \frac{{i\varepsilon h1 - im - h3}}{m} = i\left( {\cos \delta  + i\varepsilon \sin \delta } \right) = \exp i\left( {\pi /2 + \varepsilon \delta } \right),\]
and, with primary and secondary spinors being equal up to this constant phase factor, contributions (\ref{(3.4.2)}) and (\ref{(3.4.3)}) are equal. Adding the two then gives
\[ \breve {\jmath} _4^s = 2\sqrt 2 H_0^2\frac{{{{\left| m \right|}^3}}}{x}{{\mathop{\rm e}\nolimits} ^{ - 2r}},\]
and the spinor normalisation integral becomes 
\[ 1 = \int { \breve {\jmath} _4^sd\tau }  = \sqrt 2 \pi H_0^2\int\limits_{0,\infty } {dxdy\;{e^{ - 2r}}}  = \frac{{\sqrt 2 \pi H_0^2}}{{\sin \delta \left( {1 - \cos \delta } \right)}},\]
finally defining the normalisation coefficient through 
\[ H_0^2 = \frac{{\sin \delta \left( {1 - \cos \delta } \right)}}{{\sqrt 2 \pi }}.\]
The normalised moderated spinor modulus is therefore 
\[H = \frac{{{{\left| m \right|}^{3/2}}k{e^{ - r}}}}{{{2^{1/4}}\sqrt {\pi x} }},\quad k := \sqrt {\sin \delta \left( {1 - \cos \delta } \right)}, \]
with normalised spinor density 
\begin{equation}\label{(3.4.11)} \breve {\jmath} _4^s = \frac{{2{{\left| m \right|}^3}{k^2}{e^{ - 2r}}}}{{\pi x}} = \frac{{{m^2}{k^2}{e^{ - 2r}}}}{{\pi \rho }}.\end{equation}

\section{Physical Properties of Cochlean Solutions}\label{sec:physics}

\subsection{Determining Electromagnetic Potentials}\label{sec:physicspotentials}

Recalling the discussion of Section \ref{sec:poincarecurrents} that argues the acceptability of pseudoequal Poincaré electromagnetic potentials, and noting again that Poincaré and Cochlean spaces are locally identical in the asymptotic region where $m\rho >> 1$, the Poincaré form of the electromagnetic potential will be identified here with the Cochlean form
\[ {\tilde a_\mu} \equiv \breve{a} _\mu \equiv {a_\mu }\]
As indicated, this allows inclusion/omission of accents on potentials as needed, and it also requires that the potential be normalised to give the observed Poincaré charge. 

The form of the potential must be established within the Cochlean formulation, where it appears in the spinor identity as a purely geometric spinor connection that will be determined here from the spinor matter content of the Cochlean spacetime, much as the purely geometric Einstein tensor of General Relativity is determined by the matter content of the semi Riemannian spacetime.

The Cochlean electromagnetic potential – which, as noted earlier, decouples in the spinor identity itself – will be determined by insisting that it give an electromagnetic current that is a multiple of the spinor current. It will be recalled that conservation of these Cochlean currents is unaffected by overall factors depending only upon the coordinates $\rho$ and $z$ so that one may write 
\begin{equation}\label{(4.1.2)}\breve{\jmath} _\mu ^e = F\breve{\jmath} _\mu ^s \Rightarrow  - \left( {d_\rho ^2 + d_z^2} \right){\breve{a} _4} = {F_0}m\rho \breve{\jmath}_\mu ^s,\end{equation}
where the factor relating the two currents has been taken to have the form $F = F_{0}m\rho$, with a dimensionless $F_0 $ independent of spacetime coordinates. This form has been chosen for the relatively simple potentials it gives, up to a possible addition of a harmonic function in $\rho$ and $z$. 

Using the earlier moderated solutions for the spinor current, it follows that
\[ \breve{\jmath} _\mu ^e = F\breve{\jmath}  _\mu ^s \Rightarrow  - \left( {d_\rho ^2 + d_z^2} \right){\mathord{\buildrel{\lower3pt\hbox{$\scriptscriptstyle\smile$}} 
\over a} _4} = \frac{{\sin \delta \left( {1 - \cos \delta } \right)}}{\pi }{F_0}{m^3}{e^{ - \left( {x\left( {1 - \cos \delta } \right) + y\sin \delta } \right)}},\]
which gives
\[\breve{a} _4 =  - {F_0}m\frac{{\sin \delta }}{{8\pi }}{e^{ - \left( {x\left( {1 - \cos \delta } \right) + y\sin \delta } \right)}}.\]
Given earlier assumptions, this potential must be normalised in the standard Poincaré way,
\begin{eqnarray}
\tilde q &= & - \int\limits_{\scriptstyle all\atop
\scriptstyle space} {{{\vec \nabla }^2}{{\tilde a}_4}d\tau }  =  - 4\pi \int\limits_{0,\infty } {\rho d\rho dz\left\{ {\frac{1}{\rho }{d_\rho }\left( {\rho {d_\rho }{{\tilde a}_4}} \right) + d_z^2{{\tilde a}_4}} \right\}} \nonumber\\
 &= & - 4\pi \int\limits_{0,\infty } {\left[ {\rho {d_\rho }{{\tilde a}_4}} \right]_{\rho  = 0}^{\rho  = \infty }dz}  - 4\pi \int\limits_{0,\infty } {\rho \left[ {{d_z}{{\tilde a}_4}} \right]_{z = 0}^{z = \infty }d\rho } \nonumber \\
& = & - \varepsilon {m^2}{F_0}{\sin ^2}\delta \int\limits_{0,\infty } {\rho {e^{2\varepsilon m\left( {1 - \cos \delta } \right)\rho }}d\rho }  =  - \varepsilon {F_0}\frac{{{{\sin }^2}\delta }}{{4{{\left( {1 - \cos \delta } \right)}^2}}}. \nonumber
\end{eqnarray}
The resulting electromagnetic potential is then, up to harmonic terms, 
\begin{equation}\label{(4.1.6)}{\breve{a} _4} = {\tilde a_4} = \tilde q\varepsilon m\frac{{{{\left( {1 - \cos \delta } \right)}^2}}}{{2\pi \sin \delta }}{e^{ - \left( {x\left( {1 - \cos \delta } \right) + y\sin \delta } \right)}},\end{equation}
whilst the energy density for this potential is 
\[ \breve{t} _{44}^e = {\left( {{d_\rho }{a_4}} \right)^2} + {\left( {{d_z}{a_4}} \right)^2} = 2{m^4}{\tilde q^2}\frac{{{{\left( {1 - \cos \delta } \right)}^4}}}{{{\pi ^2}\left( {1 + \cos \delta } \right)}}{e^{ - 2\left( {x\left( {1 - \cos \delta } \right) + y\sin \delta } \right)}}.\]

It should be noted that the potential (\ref{(4.1.6)}) changes sign under particle conjugation: 
\begin{equation}\label{(4.1.14)}\left( {\varepsilon  \to  - \varepsilon  \wedge m \to  - m} \right) \Rightarrow \left( {\tilde q \to  - \tilde q \wedge {{\tilde a}_4} \to  - {{\tilde a}_4}} \right).\end{equation}

\subsection{The Spinor Energy-Momentum Density Tensor}

In order to calculate the properties of the states associated with the physical spinors given in Section~\ref{sec:normspindensity}
 and using the electromagnetic potentials of Section~\ref{sec:physicspotentials}, it is essential to know the total energy-momentum density tensor. Given that $\Theta_3 $  vanishes, the spinor field and its electromagnetic potential are not coupled through the spinor identity, although the spinor current remains the source of the electromagnetic field, and it is perhaps reasonable to take the total energy-momentum for the system to be simply a sum of spinor and electromagnetic contributions, without an interaction term. 
In keeping with the pragmatic approach of the current paper a form will be proposed for the spinor contribution that is quadratic in the current, 
\begin{equation}\label{(4.2.1)}{\breve t} _{\mu \nu }^s \propto {\breve {\jmath}} _{\mu} ^s {\breve {\jmath}} _{\nu} ^s.\end{equation}
Such a form is automatically conserved, since the current is independent of time and azimuth whilst the only non-zero components of (\ref{(4.2.1)}) are $ t_{22}$, $ t_{24}$,$ t_{42}$, and $ t_{44}$, and the Cochlean connection (\ref{(3.1.5)}) has distinct lower indices, of which the second is neither 2 nor 4. Furthermore, any product of a form such as this with a function of  $\rho$ and $z$ only will also be conserved. It should be noted too that from (\ref{(3.2.5)}), a spinor energy density tensor of this form will, like that for electromagnetism, be traceless.
In order to guarantee correct dimensions, as well as the convergence of integrals for small $\rho$, a factor of $\rho ^2 $ will be used in (\ref{(4.2.1)}), so that 
\begin{equation}\label{(4.2.2)}{\breve {t}} _{\mu \nu }^s = {\pi ^2}L{\rho ^2}{\breve {\jmath}} _\mu ^s{\breve {\jmath}} _\nu ^s , \end{equation}
where $L $ is a dimensionless constant.

\subsection{Total Mass and Spin}\label{sec:massandspin}

From (\ref{(4.2.2)}) with (\ref{(2.2.22)}), and (\ref{(3.2.8)})–(\ref{(3.2.10)}), both the spinor and electromagnetic energy-momentum tensors have non-zero components only for indices that are 2 or 4, and these components satisfy pseudoequality relations
\[{\breve t}_{22} = \varepsilon \rho {\breve t}_{24} = {\rho ^2}{\breve t}_{44}\]
that establish a simple connection between mass and spin integrands. Noting the covariant definition of angular momentum density as 
\[{t_{4\alpha }}{x_\beta }e_{ \bullet  \bullet \mu \nu }^{\alpha \beta  \bullet  \bullet } = {t_{42}},\quad \;\left\langle {{e_{\alpha \beta \gamma \delta }} = \sqrt { - \det {g_{\mu \nu }}} {\varepsilon _{\alpha \beta \gamma \delta }} = \rho \,{\varepsilon _{\alpha \beta \gamma \delta }} \Rightarrow e_{ \bullet  \bullet 43}^{21 \bullet  \bullet } = {\rho ^{ - 1}}} \right\rangle, \]
where $\varepsilon _{1234}$ is the totally skew symbol and values 3 and 4 have been taken for $\mu$ and $\nu$ respectively, the fermion mass and the $z $-component of angular momentum become 
\begin{equation}\label{(4.3.3)}{m_\ell} = \int\limits_{\scriptstyle all\atop
\scriptstyle space} {\breve{t}} _{44}^{total}d\tau ,\quad {\ell _z} = \int\limits_{\scriptstyle all\atop
\scriptstyle space} {\breve{t}_{24}^{total}d\tau} = \varepsilon \int\limits_{\scriptstyle all\atop
\scriptstyle space} {\rho \,\breve{t} _{44}^{total}d\tau }. \end{equation}
The total energy density that appears in these integrands is given by
\begin{equation}\label{(4.3.4)}\breve{t} _{\mu \nu }^{total} = \breve{t} _{\mu \nu }^s + \breve{t} _{\mu \nu }^e \Rightarrow \breve{t} _{44}^{total} = {\pi ^2}L{\rho ^2}\breve{
\jmath}_4^s\breve{\jmath}_4^s + {\left( {{d_1}{a_4}} \right)^2} + {\left( {{d_3}{a_4}} \right)^2 },\end{equation}
when it follows from (\ref{(3.4.11)}) and (\ref{(4.1.6)}) that
\[\breve{t} _{44}^{total} = {m^4}{k^4}{e^{ - 4r}}\Lambda ,\quad {k^2} = \sin \delta \left( {1 - \cos \delta } \right),\quad \Lambda \left( {L,\tilde q,\delta } \right) := L + 2{\tilde q^2}\frac{{1 - \cos \delta }}{{{\pi ^2}{{\left( {1 + \cos \delta } \right)}^2}}},\]
and, using Poincaré integrals as before, the fermion mass becomes
\begin{eqnarray}
{m_\ell} &= &\int\limits_{\scriptstyle all\atop
\scriptstyle space} {\breve{t}_{44}^{total}d\tau}  = 4\pi \Lambda {m^4}{k^4}\int\limits_{0,\infty } {dz\rho d\rho {e^{ - 2x\left( {1 - \cos \delta } \right) - 2y\sin \delta }}} \nonumber \\
& = & \Lambda \frac{\pi }{{16}}\left| m \right|\sin \delta \int\limits_{0,\infty } {x{e^{ - \left( {x + y} \right)}}dxdy}  = \Lambda \frac{\pi }{{16}}\left| m \right|\sin \delta ,\nonumber
\end{eqnarray}
whilst the spin is
\begin{eqnarray}\label{(4.3.7)}
{\ell _z}& = &\varepsilon \int\limits_{\scriptstyle all\atop
\scriptstyle space} {\rho \,\breve{t} _{44}^{total}d\tau }  = 4\pi \varepsilon \Lambda {m^4}{k^4}\int\limits_{0,\infty } {dz{\rho ^2}d\rho {e^{ - 2x\left( {1 - \cos \delta } \right) - 2y\sin \delta }}} \nonumber \\
& = &\varepsilon \Lambda \frac{{\pi \sin \delta }}{{64\left( {1 - \cos \delta } \right)}}\int\limits_{0,\infty } {dxdy{x^2}{e^{ - \left( {x + y} \right)}}}  = \frac{\varepsilon }{2}\frac{{{m_\ell}}}{{\left| m \right|\left( {1 - \cos \delta } \right)}}.
\end{eqnarray}
Noting the identification of ${m_\ell}$ in (\ref{(3.3.24)}), it follows that the fermion spin must be $1/2 $. 

It is of interest to consider whether it is possible for geometry alone, in the form of the spinor connection that is the electromagnetic potential, to determine all the physical properties of the state, including mass and spin. This requires setting $L $ in the above to vanish, when from (\ref{(4.3.7)}) a spin of $\varepsilon /2$ gives 
\[1 = {\tilde q^2}\frac{{\sin \delta }}{{8\pi {{\left( {1 + \cos \delta } \right)}^2}}} \Rightarrow \alpha  := \frac{{{{\tilde q}^2}}}{{4\pi }} = 2\frac{{{{\left( {1 + \cos \delta } \right)}^2}}}{{\sin \delta }} = 4{\cos ^2}\left( {\frac{\delta }{2}} \right)\cot \left( {\frac{\delta }{2}} \right).\]
This fixes $\delta $ at close to 2.8972, independently of mass, and leads to a contradiction with (\ref{(4.3.7)}) as long as $M_P $ is the underlying mass parameter. 

\subsection{Magnetic Moments of Moderated Solutions}

When cylinder coordinates are used to describe static and axially-symmetric electromagnetic fields, electrostatic multipole potentials are simply derivatives with respect to the variable $z$ of the monopole form that is $1/r$. At the same time, magnetostatic multipole potentials can be obtained by taking the derivative with respect to $\rho$ of the corresponding lower-order electrostatic form. It follows then that the multipole expansion of an axially-symmetric magnetostatic potential may be written in the form
\begin{equation}\label{(4.4.1)}\frac{{{a_\varphi }\left( {\rho ,z} \right)}}{\rho } = \sum\limits_{n = 1}^\infty  {{B_{n + 1}}{d_\rho }{\Theta _n}\left( {\rho ,z} \right)}, \end{equation}
where the $B_n $ are constants, and
\[ {\Theta _n}\left( {\rho ,z} \right) = {d_z}^{n - 1}{\Theta _1}\left( {\rho ,z} \right),\quad {\Theta _1}\left( {\rho ,z} \right) = \frac{1}{{\sqrt {{\rho ^2} + {z^2}} }}.\]
This gives the $\Theta_n$ as constant multiples of powers of $r $ with Legendre polynomials in $cos\theta $ and the magnetic dipole moment specifically as  $-4\pi B_2$. Integrating on $\rho $, (\ref{(4.4.1)}) gives
\[ \int {\frac{{{a_\varphi }\left( {\rho ,z} \right)}}{\rho }d\rho }  = \varepsilon \int {{a_4}\left( {\rho ,z} \right)d\rho }  = \sum\limits_{n = 1}^\infty  {{B_{n + 1}}{\Theta _n}\left( {\rho ,z} \right)}. \]
If this were a multipole expansion of an electrostatic potential, the first term would give the Coulomb form, with a coefficient that, up to a factor of $4\pi $, would be the total charge of the system. The same mathematics, based upon Gauss's theorem, when used here determines the dipole moment, as in
\begin{eqnarray}\label{(4.4.4)}
\frac{{{\mu _M}}}{{4\pi }}& = & - {B_2} = \frac{{ - \varepsilon }}{{4\pi }}\int\limits_{\scriptstyle all\atop
\scriptstyle space} {{{\vec \nabla }^2}\left( {\int {{a_4}\left( {\rho ,z} \right)d\rho } } \right)d\tau } \nonumber \\
 &=  &- \varepsilon \int\limits_{0,\infty } {\rho d\rho dz\left\{ {\left( {\frac{1}{\rho }{d_\rho }\left( {\rho {d_\rho }} \right) + d_z^2} \right)\left( {\int {{a_4}\left( {\rho ,z} \right)d\rho } } \right)} \right\}} \\
 &= & - \varepsilon \int\limits_{0,\infty } {\left[ {\rho {a_4}} \right]_{\rho  = 0}^{\rho  = \infty }dz}  + \varepsilon \int\limits_{0,\infty } {\rho {{\left. {\int {{d_z}{a_4}d\rho } } \right|}_{z = 0}}d\rho }  = \varepsilon \int\limits_{0,\infty } {\rho {{\left. {\int {{d_z}{a_4}d\rho } } \right|}_{z = 0}}d\rho }, \nonumber 
\end{eqnarray}
where standard assumptions of boundedness and regularity of $a_4$ have been made.
Taking  ${a_4} = \tilde q\varepsilon m\frac{{{{\left( {1 - \cos \delta } \right)}^2}}}{{2\pi \sin \delta }}{e^{ - \left( {x\left( {1 - \cos \delta } \right) + y\sin \delta } \right)}}$  from (\ref{(4.1.6)}), it follows that
 \begin{eqnarray}\label{(4.4.5)}
{\mu _M} &= &2m\tilde q\frac{{{{\left( {1 - \cos \delta } \right)}^2}}}{{\sin \delta }}\int\limits_{0,\infty } {{{\left. {\rho \left( {\int {{d_z}{e^{2\varepsilon m\left( {\rho \left( {1 - \cos \delta } \right) + z\sin \delta } \right)}}d\rho } } \right)} \right|}_{z = 0}}d\rho } \nonumber \\
 &= & 4\varepsilon {m^2}\tilde q{\left( {1 - \cos \delta } \right)^2}\int\limits_{0,\infty } {\rho \left( {\int {{e^{2\varepsilon m\rho \left( {1 - \cos \delta } \right)}}d\rho } } \right)d\rho }  = \frac{{\tilde q}}{{2m\left( {1 - \cos \delta } \right)}}\int\limits_{0,\infty } {x{e^{ - x}}dx}\nonumber \\
 &= & \frac{{ - \varepsilon \tilde q}}{{2{m_\ell}}} ,
\end{eqnarray}
showing the moment to be exactly one magneton. It should be noted that although the sign of $\mu _M $ is unchanged under the particle conjugation of (\ref{(4.1.14)}), the spin in (\ref{(4.3.7)}) reverses, so the relation between moment and spin changes sign as required. 

The result (\ref{(4.4.5)}) is typical of an essentially non-interacting elementary Dirac particle and is normally derived in introductory texts such as \cite{bjorkendrell} on largely algebraic grounds. The derivation given here, based as it is on analytic properties of an axial current, suggests that the charge distribution of a simple moderated solution to the Cochlean form of the spinor identity (\ref{(2.1.7)}) characterises such an elementary particle. 

\section{Summary and Outlook}\label{sec:summary}

In this paper a classical geometrically-based treatment accounts for what is normally understood as the intrinsic spin exhibited by Dirac particles. It does so in a way that associates an extended spacetime structure given by torsion with the otherwise strictly localised particle. This reliance upon geometry produces an analytic, transitionally quantal approach to a spinning charge density that leads to the same magnetic moment as that resulting from Dirac’s original quantum-mechanical treatment of charged fermions. 

In order fully to describe real particles however, and especially to calculate their masses, it is expected that extension of the current flat spacetime formulation to more general semi-Riemannian spacetimes will be necessary, and further calculation is therefore deferred until that has been presented in a second essentially formulational paper.

\end{document}